\documentclass[preprint,superscriptaddress,amsmath,amssymb,aps,prd,nofootinbib]{revtex4-1}

\usepackage{graphicx}
\usepackage{dcolumn}
\usepackage{bm}

\usepackage{amsmath}
\usepackage{bm}
\usepackage{slashed}
\usepackage{epsfig}
\usepackage{amsfonts}
\usepackage{epstopdf}
\usepackage{color}
\usepackage{float}

\begin{document}


\title{Next-to-leading-order relativistic and QCD corrections to prompt $\boldsymbol{J/\psi}$ pair photoproduction at future $\boldsymbol{e^+e^-}$ colliders}

\author{Zhi-Guo He}
\affiliation{Department of Physics and Electronics,
School of Mathematics and Physics,
Beijing University of Chemical Technology, Beijing 100029, China}
\affiliation{{II.} Institut f\"ur Theoretische Physik, Universit\"at Hamburg,
Luruper Chaussee 149, 22761 Hamburg, Germany}
\affiliation{Institut f\"ur Theoretische Physik, 
Universit\"at Regensburg, 93040 Regensburg, Germany}
\author{Xiao-Bo Jin}%
\affiliation{Center of Advanced Quantum Studies, Department of Physics, Beijing Normal University, Beijing 100875, China}
\author{Bernd A. Kniehl}
\affiliation{{II.} Institut f\"ur Theoretische Physik, Universit\"at Hamburg,
Luruper Chaussee 149, 22761 Hamburg, Germany}
\author{Rong Li}%
\affiliation{MOE Key Laboratory for Nonequilibrium Synthesis and Modulation of Condensed Matter, School of Physics, Xi'an Jiaotong University, Xi'an 710049, China} 
\affiliation{ Institute of Theoretical Physics, Xi'an Jiaotong University, Xi'an 710049, China}

\date{\today}

\begin{abstract}
Within the framework of nonrelativistic-QCD factorization, we calculate both the 
next-to-leading-order relativistic and QCD corrections to prompt $J/\psi$ pair production, with 
feeddown from $\psi(2S)$ mesons, via photon-photon collisions at future $e^+e^-$ colliders including 
the Future Circular Lepton Collider (FCC-ee), the Circular Electron Positron Collider (CEPC), and 
the Compact Linear Collider (CLIC). We present total cross sections and distributions in single 
$J/\psi$ transverse momentum and rapidity, and in $J/\psi$ pair invariant mass. The relativistic and 
QCD corrections both turn out to be large and negative. Yet, the production rates are large enough 
for useful experimental studies.
\end{abstract}

\maketitle


\section{Introduction}\label{sec:intro}

The production of heavy quarkonia, including the $J/\psi$ meson as the most prominent specimen, provides a 
perfect laboratory to explore the interplay between perturbative and nonperturbative phenomena in quantum 
chromodynamics (QCD), since it accommodates the creation of a heavy quark pair $Q\bar{Q}$ at high energy 
as well as its transition into a heavy meson at low energy. The framework of nonrelativistic-QCD (NRQCD) 
factorization~\cite{Bodwin:1994jh} has been very successful in explaining the heavy-quarkonium production 
mechanism. However, the $J/\psi$ polarization puzzle \cite{Butenschoen:2012px} and other problems in the validation of the predicted 
universality of the NRQCD long-distance matrix elements (LDMEs) \cite{Butenschoen:2014dra,Butenschoen:2022wld} still challenge NRQCD factorization.
Shortly after the birth of NRQCD factorization, prompt $J/\psi$ pair hadroproduction was proposed as a 
showcase for the color-octet (CO) mechanism, a key feature of NRQCD factorization, because the 
hadronization occurs there twice~\cite{Barger:1995vx}. Later on, the color-singlet (CS) channel was found 
to contribute predominately in the region of small and moderate $J/\psi$ transverse momentum, $p_T^{J/\psi}
$~\cite{Qiao:2002rh,Li:2009ug,Ko:2010xy,He:2015qya}. Meanwhile, prompt $J/\psi$ pair hadroproduction is 
also viewed as a good probe of the double parton scattering (DPS) mechanism in hadron collisions and as a 
tool to extract its key parameter $\sigma_{\mathrm{eff}}$~\cite{Kom:2011bd}. However, fit results of 
$\sigma_\mathrm{eff}$ from experimental~\cite{D0:2014vql,LHCb:2016wuo,ATLAS:2016ydt, LHCb:2023ybt} and 
theoretical~\cite{Lansberg:2014swa,Prokhorov:2020owf} analyses exhibit an incoherent picture.

Since the discovery of the Higgs boson at the CERN LHC~\cite{ATLAS:2012yve,CMS:2012qbp}, to 
build next-generation $e^{+}e^{-}$ colliders reaching center-of-mass energies of hundreds of 
GeV or even a few TeV for high-precision studies of the electroweak sector of the standard 
model has been on the agenda of the high-energy physics community. Possible realizations 
include the CERN Future Circular Lepton Collider (FCC-ee)~\cite{FCC:2018evy}, the Circular 
Electron Positron Collider (CEPC)~\cite{CEPCStudyGroup:2018rmc} in China, and the CERN Compact 
Linear Collider (CLIC)~\cite{CLICdp:2018cto}. As we will argue below, such high-luminosity 
$e^{+}e^{-}$ colliders will also offer great opportunities to deepen our understanding of 
the double prompt $J/\psi$ production mechanism. A decisive advantage of $e^{+}e^{-}$ colliders 
versus hadron colliders resides in the absence of DPS. Historically, inclusive single $J/\psi$ 
production in two-photon collisions at the CERN Large Electron Positron Collider (LEP) is among 
the earliest evidences of the CO mechanism~\cite{Klasen:2001cu,Butenschoen:2011yh}.

In $e^+e^-$ collisions, there are generally two distinct production modes, $e^+e^-$ 
annihilation and two-photon scattering, where the photons originate from both electromagnetic 
bremsstrahlung and synchrotron radiation off the colliding bunches known as beamstrahlung.
The photons can either directly participate in the hard collision as pointlike particles, or as 
resolved photons via their quark and gluon contents as described by photonic parton density 
functions (PDFs)~\cite{DeWitt:1978wn}. This results in three production channels: direct, 
single resolved, and double resolved. $J/\psi$ pair production by $e^{+}e^{-}$ annihilation was 
investigated by several 
groups~\cite{Bodwin:2002fk,Bodwin:2002kk,Hagiwara:2003cw,Bodwin:2006yd,Braguta:2007ge,Gong:2008ce,Fan:2012dy}, 
even through next-to-next-to-leading order in the strong-coupling constant 
$\alpha_s$~\cite{Sang:2023liy,Huang:2023pmn}. $J/\psi$ pair production by two-photon scattering 
was first considered by Qiao in 2002 as a contribution to inclusive $J/\psi$ 
production~\cite{Qiao:2001wv}. Recently, exclusive $J/\psi$ pair production by two-photon 
scattering at $e^{+}e^{-}$ colliders was studied through next-to-leading order (NLO) in 
$\alpha_s$, with the result that QCD corrections, of relative order $\mathcal{O}(\alpha_s)$, 
can decrease theoretical predictions by almost $80\%$~\cite{Yang:2020xkl}.

In the charmonium rest frame, the charm quark relative velocity $v$ is not small, with 
$v^2\sim\alpha_s(2m_c)$, which explains why relativistic corrections, being of relative order 
$\mathcal{O}(v^2)$, may be comparable to QCD corrections. In fact, in $e^+e^-$ annihilation at 
center-of-mass energy $\sqrt{S}=10.6$~GeV, $\mathcal{O}(v^2)$ corrections largely enhance the LO predictions 
for exclusive double charmonium~\cite{He:2007te} and inclusive 
$J/\psi+X_\mathrm{non-c\bar{c}}$~\cite{He:2009uf,Jia:2009np} production, and they are also considerable in 
$J/\psi$ photo- and hadroproduction, for both yield~\cite{Xu:2012am,He:2014sga} and 
polarization~\cite{He:2015gla}. As for prompt $J/\psi$ pair hadroproduction, the 
$\mathcal{O}(v^2)$ corrections to the CS channel were found to reduce the cross section appreciably in the 
large-$p_T$ region~\cite{Li:2013csa} and substantially near threshold~\cite{He:2024ugx}.
This suggests that $\mathcal{O}(v^2)$ corrections may also be important in the case of $J/\psi$ pair 
production in two-photon scattering, the more so as $\mathcal{O}(\alpha_s)$ corrections were found to 
dramatically reduce the cross section~\cite{Yang:2020xkl}, as already mentioned above.
This strongly motivates our analysis below. Since an independent cross check of the results of 
Ref.~\cite{Yang:2020xkl} is still lacking, we also recalculate the $\mathcal{O}(\alpha_s)$ corrections 
here. 

The rest of this paper is organized as follows. In Sec.~\ref{sec:framework}, we describe techniques to 
calculate the $\mathcal{O}(v^2)$ and $\mathcal{O}(\alpha_s)$ corrections to the short-distance 
coefficients (SDCs) within the NRQCD factorization framework. In Sec.~\ref{sec:ph}, we present numerical 
predictions appropriate for FCC-ee, CEPC, and CLIC experimental conditions as anticipated today.
In Sec.~\ref{sec:summary}, we summarize our conclusion.

\section{Theoretical framework}\label{sec:framework}

According to the factorization theorem of the QCD parton model, the differential cross section of inclusive prompt $J/\psi$ pair production by two-photon scattering may be expressed as (see, e.g., Ref.~\cite{Klasen:2001cu})
\begin{eqnarray}
&&\mathrm{d}\sigma(e^{+}e^{-}\to e^{+}e^{-}+2J/\psi+X)
  =\sum_{i,j,H_1,H_2}
  \int\mathrm{d}x_1\mathrm{d}x_2
  f_{\gamma}(x_1)f_{\gamma}(x_2)
  \int d x_i d x_j f_{i/\gamma}(x_i)f_{j/\gamma}(x_j)
  \nonumber  \\
&&{}\times  \mathrm{d}\hat{\sigma}(i+j\to H_1+H_2)
  \mathrm{Br}(H_1 \rightarrow J/\psi+X)
  \mathrm{Br}(H_2 \rightarrow J/\psi+X)\,,
\label{eq:gamma gamma}
\end{eqnarray}
where $f_{\gamma}(x)$ is the scaled-energy distribution of the photons from both bremsstrahlung and beamstrahlung off the incoming leptons, $f_{i/\gamma}(x)$ is the PDF in longitudinal-momentum fraction of parton $i$ in the resolved photon, being $f_{\gamma/\gamma}(x)=\delta(1-x)$ for the direct photon, $\mathrm{d}\hat{\sigma} (i+j\to H_1+H_2)$ is the partonic differential cross section for associated production of charmonium states $H_1,H_2=J/\psi,\chi_{cJ},\psi(2S)$, and $\mathrm{Br}(H \rightarrow J/\psi+X)$ is the branching fraction of $H$ decay into $J/\psi$, being 1 for $H=J/\psi$ implying direct production. In Eq.~\eqref{eq:gamma gamma}, $X$ collectively denotes the undetected particles in the respective final states, reflecting the inclusiveness of the experimental observation mode.

At parton level, the Feynman diagrams of $2J/\psi$ production are the same as those of $J/\psi+c\bar{c}$ 
production, which was studied for two-photon scattering in Ref.~\cite{Li:2009zzu}, where the single-resolved and double-resolved contributions were found to be greatly suppressed.
We recover this finding at LO for direct $2J/\psi$ production in two-photon scattering under FCC-ee experimental conditions, where we find the single-resolved and double-resolved contributions to be more than 3 orders of magnitude smaller than the direct contribution.
In the following, we thus concentrate on direct photoproduction. In the latter case, CO contributions are known to be important only at large values of $p_T^{J/\psi}$~\cite{He:2015qya}, where the cross sections are likely to be 
too small to be measurable in the first few years of running at the above-mentioned future $e^+e^-$ 
colliders. In the following, we thus focus on CS contributions, which arise from the partonic subprocesses
\begin{eqnarray}
\gamma+\gamma&\to& (c\bar{c})_1({}^3S_1^{[1]})+(c\bar{c})_2({}^3S_1^{[1]})\,,
  \label{eq:ss}\\
\gamma+\gamma&\to& (c\bar{c})_1({}^3P_{J_1}^{[1]})+(c\bar{c})_2({}^3P_{J_2}^{[1]})\,,
\end{eqnarray}
yielding $2J/\psi$, $J/\psi+\psi(2S)$, $2\psi(2S)$, and $\chi_{c_{J1}}+\chi_{c_{J2}}$ final states.
We verified that, under FCC-ee experimental conditions, the production rates of the $\chi_{c_{J1}}+\chi_{c_{J2}}$ channels are about one order of magnitude smaller than those of the other channels, and they are reduced by two additional factors of branching fraction to become negligibly small feed-down contributions.
We thus concentrate on partonic subprocess~(\ref{eq:ss}) and compute relativistic corrections of $\mathcal{O}(v^2)$ and QCD corrections of $\mathcal{O}(\alpha_s)$ to its cross section.
By color conservation, the latter are purely virtual.

Through $\mathcal{O}(v^2)$ in NRQCD, the relevant partonic cross section appearing in Eq.~(\ref{eq:gamma gamma}) factorizes as
\begin{eqnarray}
\lefteqn{\mathrm{d}\hat{\sigma}(\gamma+\gamma\to H_1+H_2)
  =\sum_{m,n,H_1,H_2} 
  \left(\frac{d F(m,n)}{m_c^{d_{\mathcal{O}(m)}-4}m_c^{d_{\mathcal{O}(n)}-4}}
  \langle \mathcal{O}^{H_1}(m)\rangle \langle \mathcal{O}^{H_2}(n)\rangle
  \right.}
\nonumber\\
&&{}+\left.\frac{d G_{1}(m,n)}{m_c^{d_{\mathcal{P}(m)}-4}m_c^{d_{\mathcal{O}(n)}-4}}
  \langle \mathcal{P}^{H_1}(m)\rangle \langle \mathcal{O}^{H_2}(n)\rangle
  +\frac{d G_{2}(m,n)}{m_c^{d_{\mathcal{O}(m)}-4}m_c^{d_{\mathcal{P}(n)}-4}}
  \langle \mathcal{O}^{H_1}(m)\rangle \langle \mathcal{P}^{H_2}(n)\rangle
  \right)\,,
\label{eq:nrqcd}
\end{eqnarray}
where $m,n={}^{2S+1}L_J^{[a]}$ are quantum numbers in spectroscopic notation, with total spin $S$, orbital angular momentum $L$, total angular momentum $J$, and color configuration $a=1,8$ for CS and CO;
$\mathcal{O}^{H}(n)$ and $\mathcal{P}^{H}(n)$ are four-fermion operators of mass dimensions $d_{\mathcal{O}}$ and $d_{\mathcal{P}}$ describing the nonperturbative transition $n\to H$ at LO and $\mathcal{O}(v^2)$;
$\langle\mathcal{O}^{H}(n)\rangle$ and $\langle\mathcal{P}^{H}(n)\rangle$ are the respective LDMEs; and
$F(m,n)$ and $G_i(m,n)$ are the appropriate SDCs.
Definitions of the relevant four-fermion operators,
$\mathcal{O}^H({}^3S_1^{[1]})$ and $\mathcal{P}^H({}^3S_1^{[1]})$, may be found, e.g., in Eq.~(3) of Ref.~\cite{He:2024ugx}.
The calculation of $F({}^3S_1^{[1]},{}^3S_1^{[1]})$ and $G_i({}^3S_1^{[1]},{}^3S_1^{[1]})$ proceeds in a fashion similar to the hadron collider case in Ref.~\cite{He:2024ugx} and references cited therein, and we merely list our final results.
Notice that, according to the $p_T$ power counting rules of Ref.~\cite{He:2015qya}, we have $\mathrm{d}\hat{\sigma}/\mathrm{d}p_T^2 \propto 1/p_T^8$ for partonic subprocess~\eqref{eq:ss}.

Starting from the Mandelstam variables $s=(k_1+k_2)^2$, $t=(k_1-P_1)^2$, and $u=(k_1-P_2)^2$ of process $\gamma(k_1)+\gamma(k_2)\to H_1(P_1)+H_2(P_2)$, we denote the counterparts of $t$ and $u$ in the nonrelativistic limit as $t_0$ and $u_0$, and define $\hat{t}_0=t_0+(s-8m_c^2)/2$ and $\hat{u}_0=u_0+(s-8m_c^2)/2$.
The two-body phase space element in the nonrelativistic limit may thus be written as
\begin{equation}
d\Phi_{20}= \frac{\mathrm{d}\hat{t}_0 \mathrm{d}\hat{u}_0} {8\pi s}
\delta\left(\hat{t}_0+\hat{u}_0\right)\,,
\end{equation}
and receives the $\mathcal{O}(v^2)$ correction factor $K=-4/(s-16m_c^2)$.
Our final results then read
\begin{eqnarray}
\frac{F({^3S_1^{[1]},^3S_1^{[1]}})}{m_c^4}&=&\frac{1}{2s}\int d\Phi_{20} \left(|M_0|^2+|M_1|^2\right)\,, 
\nonumber\\
\frac{G_{1}({^3S_1^{[1]},^3S_1^{[1]}})}{m_c^6} &=&\frac{G_{2}({^3S_1^{[1]},^3S_1^{[1]}})}{m_c^6} =
\frac{1}{2s}\int d\Phi_{20} \left(K|M_0|^2+|N|^2\right)\,,
\end{eqnarray}
where $|M_0|^2$ is the absolute square of the tree-level amplitude, $|M_1|^2$ stands for its $\mathcal{O}(\alpha_s)$ correction, and $|N|^2$ for its $\mathcal{O}(v^2)$ correction.
The expressions for $|M_0|^2$, $|M_1|^2$, and $|N|^2$ assume relatively compact forms when they are written in terms of $s$ and $\hat{t}_0$.

Below, we list $|M_0|^2$ for reference and $|N|^2$ as a new result:
\begin{eqnarray}
|M_0|^2
&=&
\frac{262144 \pi ^4 \alpha ^2 e_c^4 \alpha _s^2}{729 m_c^2 \left(s^3-4 s \hat{t}_0^2\right)^4}
\left[
s^{10}
+48 s^9 m_c^2
-16 s^8 \left(\hat{t}_0^2-188 m_c^4\right)
-512 s^7 m_c^2 \left(4 m_c^4+\hat{t}_0^2\right)
\right.
\nonumber\\
&&{}+96 s^6\left(128 \hat{t}_0^2 m_c^4+1024 m_c^8+\hat{t}_0^4\right)
+1536 s^5 \hat{t}_0^2 m_c^2\left(\hat{t}_0^2-16 m_c^4\right)
\nonumber\\
&&{}-256 s^4 \left(-3072 \hat{t}_0^2 m_c^8-88 \hat{t}_0^4m_c^4+\hat{t}_0^6\right)
+32768 s^3 \hat{t}_0^4 m_c^6
\nonumber\\
&&{}+\left.256 s^2 \left(128 \hat{t}_0^6 m_c^4
+6144\hat{t}_0^4 m_c^8+\hat{t}_0^8\right)
-4096 s \hat{t}_0^6 m_c^2 \left(\hat{t}_0^2
-96m_c^4\right)+49152 \hat{t}_0^8 m_c^4
\right]\,,\quad
\label{eq:born}\\
|N|^2
&=&
-\frac{131072 \pi ^4 \alpha ^2 e_c^4 \alpha _s^2}{2187 m_c^4 \left(s^3-4 s \hat{t}_0^2\right)^5\left(s-16 m_c^2\right)}
\left[3 s^{14}-144 s^{13} m_c^2
+s^{12} \left(832 m_c^4-76 \hat{t}_0^2\right)
\right.
\nonumber\\
&&{}+64 s^{11} \left(55 \hat{t}_0^2 m_c^2-2368
m_c^6\right)+32 s^{10} \left(-1192 \hat{t}_0^2 m_c^4+82432 m_c^8+25 \hat{t}_0^4\right)
\nonumber\\
&&{}-1024 s^9
m_c^2 \left(-2424 \hat{t}_0^2 m_c^4+14336 m_c^8+27 \hat{t}_0^4\right)
\nonumber\\
&&{}-128 s^8 \left(-29184
\hat{t}_0^2 m_c^8-1488 \hat{t}_0^4 m_c^4-1769472 m_c^{12}+35 \hat{t}_0^6\right)
\nonumber\\
&&{}+2048 s^7
\hat{t}_0^2 m_c^2 \left(2656 \hat{t}_0^2 m_c^4+50176 m_c^8+47 \hat{t}_0^4\right)
\nonumber\\
&&{}+256 s^6
\left(4325376 \hat{t}_0^2 m_c^{12}+98304 \hat{t}_0^4 m_c^8-1184 \hat{t}_0^6 m_c^4+55
\hat{t}_0^8\right)
\nonumber\\
&&{}-4096 s^5 \hat{t}_0^4 m_c^2 \left(-4352 \hat{t}_0^2 m_c^4-100352 m_c^8+49
\hat{t}_0^4\right)
\nonumber\\
&&{}-1024 s^4 \hat{t}_0^4 \left(-204800 \hat{t}_0^2 m_c^8-272 \hat{t}_0^4
m_c^4-2752512 m_c^{12}+23 \hat{t}_0^6\right)
\nonumber\\
&&{}+16384 s^3 \hat{t}_0^6 m_c^2 \left(-384 \hat{t}_0^2
m_c^4-75776 m_c^8+27 \hat{t}_0^4\right)
\nonumber\\
&&{}+16384 s^2 \hat{t}_0^6 \left(-4352 \hat{t}_0^2 m_c^8-68
\hat{t}_0^4 m_c^4+491520 m_c^{12}+\hat{t}_0^6\right)
\nonumber\\
&&{}-\left.131072 s \hat{t}_0^8
m_c^2 \left(112 \hat{t}_0^2 m_c^4-15360 m_c^8+5 \hat{t}_0^4\right)
+6291456 \hat{t}_0^{10} m_c^4
\left(40 m_c^4+\hat{t}_0^2\right)
\right]\,,
\label{eq:rel}
\end{eqnarray}
where $\alpha$ is Sommerfeld's fine-structure constant and $e_c=2/3$ is the fractional electric charge of the charm quark.  

In the computation of $|M_1|^2$, we encounter both ultraviolet (UV) and infrared (IR) divergences, which are both regularized using dimensional regularization with $D=4-2\epsilon$ space-time dimensions.
The UV divergences are removed by renormalizing the parameters and external fields of the tree-level amplitude $M_0$.
As is usually done in loop computations of heavy-quarkonium production, we adopt the on-mass-shell (OM) scheme for the renormalization of the charm quark wave function ($Z_2$) and mass ($Z_m$), and the gluon wave function ($Z_3$), and the modified minimal-subtraction ($\overline{\mathrm{MS}}$) scheme for the renormalization of the strong-coupling constant ($Z_g$).
For the reader's convenience, we list the respective one-loop expressions here:
\begin{eqnarray}
\delta Z_2^{\mathrm{OS}}
&=&-C_F\frac{\alpha_s}{4\pi}
N_\epsilon
\left(
  \frac{1}{\epsilon_{\mathrm{UV}}}
  +\frac{2}{\epsilon_{\mathrm{IR}}}
  +4
\right)+\mathcal{O}(\alpha_s^2)\,,\nonumber\\
\delta Z_m^{\mathrm{OS}}
&=&
-3C_F\frac{\alpha_s}{4\pi}
N_\epsilon
\left(
  \frac{1}{\epsilon_{\mathrm{UV}}}
  +\frac{4}{3}
\right)+\mathcal{O}(\alpha_s^2)\,,\nonumber\\
\delta Z_3^{\mathrm{OS}}
&=&
\frac{\alpha_s}{4\pi}
N_\epsilon
\left[(\beta_0^\prime-2C_A)
\left(
  \frac{1}{\epsilon_{\mathrm{UV}}}
  -\frac{1}{\epsilon_{\mathrm{IR}}}
\right)-
\frac{4}{3}
T_F(n_f-n_l)
\frac{1}{\epsilon_{\mathrm{UV}}}\right]+\mathcal{O}(\alpha_s^2)\,,\nonumber\\
\delta Z_g^{\overline{\mathrm{MS}}}
&=&
-\frac{\beta_0}{2}\,
\frac{\alpha_s}{4\pi}
N_\epsilon
\left(
  \frac{1}{\epsilon_{\mathrm{UV}}}
  +\ln\frac{m_c^2}{\mu_r^2}
\right)+\mathcal{O}(\alpha_s^2)\,,
\label{eq:ren}
\end{eqnarray}
where $N_\epsilon=(4\pi\mu_r^2/m_c^2)^\epsilon/\Gamma(1-\epsilon)$, $\mu_r$ is the renormalization scale, $\beta_0=11C_A/3-4T_Fn_f/3$ is the one-loop coefficient of the QCD beta function, $C_A=3$, $T_F=1/2$, $n_f=4$ is the number of active quark flavors, and $\beta_0^\prime$ emerges from $\beta_0$ by replacing $n_f$ with the number of light quark flavors $n_l=3$.  
Thanks to the absence of real gluon radiation, the IR singularities in $|M_1|^2$ cancel among themselves in combination with the IR divergences from wave function renormalization in Eq.~\eqref{eq:ren}.
Our result for $|M_1|^2$ is too lengthy to be listed here.
We find numerical agreement with the results for double charmonium photoproduction in Ref.~\cite{Yang:2020xkl} upon adopting the inputs specified there.
Looking at the one-loop diagrams in Fig.~2 of Ref.~\cite{Yang:2020xkl}, we note that all of them scale with $e_c^2$, except for the box diagrams in Fig.~2(l), which scale with $e_q^2$.
Keeping this in mind, our analytic results readily carry over to double bottomonium production.\footnote{%
We can also reproduce the numerical results for double bottomonium production in Ref.~\cite{Yang:2020xkl}, if we attach the superfluous factor of $e_b^2/e_c^2=1/4$ to the contribution due to the box diagrams mentioned above.}

We use the program packages FeynArts~\cite{Kublbeck:1990xc} and QGRAF~\cite{Nogueira:1991ex} to generate Feynman diagrams and amplitudes, and FeynCalc \cite{Mertig:1990an} and FORM~\cite{Vermaseren:2000nd} to handle the Dirac and SU(3)${}_c$ algebras.
We reduce the one-loop scalar integrals to a small set of master integrals using the Laporta algorithm~\cite{Laporta:2000dsw} of integration by parts~\cite{Chetyrkin:1981qh} as implemented in the program packages Reduze~2~\cite{vonManteuffel:2012np} and FIRE6~\cite{Smirnov:2019qkx}.
We evaluate the master integrals analytically using the program packages Package-X~2.0~\cite{Patel:2016fam} and QCDloop~\cite{Ellis:2007qk}. 

\section{Numerical results}\label{sec:ph}

In the numerical analysis, we evaluate $\alpha_s^{(n_f)}(\mu_r)$ with $n_f=4$ and
$\Lambda_\mathrm{QCD}^{(4)}=215~\mathrm{MeV}$ (326~MeV) at tree level (one loop) \cite{Pumplin:2002vw}.
We set $\mu_r=\xi\sqrt{s}$, take $\xi=1$ as default, and vary $\xi$ between 1/2 and 2 to estimate the theoretical uncertainties from unknown higher orders in $\alpha_s$.
For definiteness, we choose $m_c=1.5~\mathrm{GeV}$ as is frequently done in similar analyses, so as to facilitate comparisons with the literature. 
The values $m_{J/\psi}=3.097~\mathrm{GeV}$, $m_{\psi(2S)}=3.686~\mathrm{GeV}$, $\mathrm{Br}(\psi(2S) \rightarrow J/\psi+X)=61.4\%$ for masses and branching fraction are taken from the latest Review of Particle Physics \cite{ParticleDataGroup:2022pth}.
As for the $J/\psi$ and $\psi(2S)$ CS LDMEs, we use the ones evaluated from the wave functions at the origin for the Buchm\"{u}ller-Tye potential \cite{Eichten:1995ch},
\begin{eqnarray}
  \langle \mathcal{O}^{J/\psi}({}^3S_1^{[1]})\rangle
  &=&1.16~\mathrm{GeV}^3\,,\nonumber
  \\
  \langle \mathcal{O}^{\psi(2S)}({}^3S_1^{[1]})\rangle
  &=&0.758~\mathrm{GeV}^3\,,
\end{eqnarray}
and estimate the $\mathcal{O}(v^2)$ ones by the ratio obtained for the $J/\psi$ case in Ref.~\cite{Bodwin:2006dn},
\begin{equation}\label{eq: v^2 LDMEs}
\frac{\langle \mathcal{P}^{J/\psi}({}^3S_1^{[1]})\rangle}
{\langle \mathcal{O}^{J/\psi}({}^3S_1^{[1]} \rangle}  =  
0.5~\mathrm{GeV}^2\approx  
\frac{\langle \mathcal{P}^{\psi(2S)}({}^3S_1^{[1]})\rangle}
{\langle\mathcal{O}^{\psi(2S)}({}^3S_1^{[1]})\rangle}\,.
\end{equation}
In the evaluation of feed-down contributions to differential cross sections, we
approximate the momentum of the $J/\psi$ meson from $\psi(2S)$ decay as 
\begin{equation}
p_{J/\psi} = \frac{m_{J/\psi}}{m_{\psi(2S)}}p_{\psi(2S)}\,.
\end{equation}

We include both bremsstrahlung and beamstrahlung by superimposing their spectra.
The bremsstrahlung distribution is described in the Weizsacker-Williams approximation (WWA) as \cite{Frixione:1993yw}
\begin{equation}
  f^{\mathrm{WWA}}_{\gamma}(x)
  =
  \frac{\alpha}{2\pi}
  \left[
    \frac{1+(1-x)^2}{x}
    \ln\frac{Q_\mathrm{max}^2}{Q_\mathrm{min}^2}
    +
    2m_e^2x
    \left(
      \frac{1}{Q_\mathrm{max}^2}-\frac{1}{Q_\mathrm{min}^2}
    \right)
  \right]\,,
\end{equation}
with photon virtuality $Q^2$ bounded by
\begin{eqnarray}
    Q_\mathrm{min}^2
    &=&
    \frac{m_e^2x^2}{1-x}\,,
\nonumber\\
    Q_\mathrm{max}^2
    &=&E_e^2\theta_c^2(1-x)
    +Q_\mathrm{min}^2\,,
\end{eqnarray}
where $x=E_\gamma/E_e$, $E_e=\sqrt{S}/2$ is incoming-lepton energy, $E_\gamma$ is the radiated-photon energy, and $\theta_c$ the maximum angle by which the photon is deflected from the flight direction of the emitting lepton in the center-of-mass frame.
The $x$ distribution of beamstrahlung is characterized by the effective beamstrahlung parameter $\Upsilon$.
If $\Upsilon\alt5$, a useful and convenient approximation is given by \cite{Chen:1991wd}
\begin{eqnarray}
f^{\mathrm{beam}}_{\gamma}(x)&=&\frac{1}{\Gamma(1/3)}\left(\frac{2}{3\Upsilon}\right)^{1/3}x^{-2/3}(1-x)^{-1/3}e^{-2x/[3\Upsilon(1-x)]}\nonumber\\
&&{}\times\left\{\frac{1-\sqrt{\Upsilon/24}}{g(x)}\left[1-\frac{1}{g(x)N_\gamma}\right]\left(1-e^{-g(x)N_{\gamma}}\right)+\sqrt{\frac{\Upsilon}{24}}\left[1-\frac{1}{N_\gamma}\left(1-e^{-N_\gamma}\right)\right]\right\}\,,
\nonumber\\
&&
\end{eqnarray}
where 
\begin{equation}
g(x)=1-\frac{1}{2}\left[(1+x)\sqrt{1+\Upsilon^{2/3}}+1-x \right](1-x)^{2/3}\,,
\end{equation}
and 
\begin{equation}
N_\gamma=\frac{5\alpha\sigma_zm_e^2\Upsilon}{2E_e\sqrt{1+\Upsilon^{2/3}}}
\end{equation}
is the average number of photons emitted by an electron or position during the collision, with $\sigma_z$ being the longitudinal bunch length.

\begin{table}
\caption{\label{Tab:colliders} Parameters of FCC-ee, CEPC, and CLIC relevant for our calculations.}
	\begin{ruledtabular}
	\begin{tabular}{lccccc}
		facility & $\sqrt{S}$ [GeV] & $\theta_c$ [mrad] & average $\Upsilon$ & $\sigma_z$ [mm] & $\int\mathrm{d}t\,\mathcal{L}$ [$\mathrm{ab}^{-1}$]\\
		\hline
        FCC-ee & 92 & 30 &$10^{-4}$ & 15.5 & 17\\
		CEPC & 92 & 33 & $2\times10^{-4}$ & 8.7 & 15\\  
        CLIC & 3,000 & 20 & 5& 0.044 & 0.6\\ 
	\end{tabular}
  \end{ruledtabular}
\end{table}

As mentioned in Sec.~\ref{sec:intro}, we assess here three future realizations of high-energy, 
high-luminosity $e^+e^-$ colliders under discussion by the world-wide particle physics 
community, namely FCC-ee~\cite{FCC:2018evy}, CEPC~\cite{CEPCStudyGroup:2018rmc}, and 
CLIC~\cite{CLICdp:2018cto}, with regard to their potentials to allow for measurements of prompt 
$J/\psi$ pair photoproduction. For each of them, the set of parameters relevant for our 
numerical analysis, including $e^+e^-$ center-of-mass energy $\sqrt{S}$, upper cut $\theta_c$ 
on bremsstrahlung deflection angle, average beamstrahlung parameter $\Upsilon$, longitudinal bunch 
length $\sigma_z$, and estimated luminosity per experiment $\int\mathrm{d}t\,\mathcal{L}$ integrated over one year of running, as quoted in Ref.~\cite{ParticleDataGroup:2022pth}, are collected in Table~\ref{Tab:colliders}.
Throughout our study, we impose the cut $p_T^{J/\psi}\ge2~\mathrm{GeV}$ on the transverse momentum of each $J/\psi$ meson.

\begin{table}
  \caption{\label{tab:tot}%
    NRQCD predictions of $\sigma(e^+e^-\to 2J/\psi+X)$ [fb] via photoproduction at LO and with $\mathcal{O}(v^2)$ and $\mathcal{O}(\alpha_s)$ corrections consecutively added at FCC-ee, CEPC, and CLIC with theoretical uncertainties.}
	\begin{ruledtabular}
	\begin{tabular}{lccc}
		order & FCC-ee & CEPC & CLIC\\
		\hline
                LO & $5.88^{+2.99}_{-1.70}$ & $6.00^{+3.06}_{-1.73}$ & $144^{+73}_{-36}$\\
                plus $\mathcal{O}(v^2)$ & $5.17^{+2.64}_{-1.49}$ & $5.28^{+2.69}_{-1.53}$ & $126^{+64}_{-32}$\\
                plus $\mathcal{O}(\alpha_s)$ & $2.65^{+0.34}_{-1.99}$ & $2.71^{+0.34}_{-2.03}$ & $64.6^{+7.5}_{-47.9}$\\
	\end{tabular}
  \end{ruledtabular}
\end{table}

We start by considering the total cross section $\sigma(e^+e^-\to 2J/\psi+X)$ via photoproduction.
Starting from the LO NRQCD predictions, we in turn add the $\mathcal{O}(v^2)$ and $\mathcal{O}(\alpha_s)$ corrections.
Our results for FCC-ee, CEPC, and CLIC are displayed in Table~\ref{tab:tot}, where the central values refer to $\xi=1$ and the theoretical error bands to $\xi=1/2,2$.
In all three cases, we observe that the $\mathcal{O}(v^2)$ corrections induce a reduction by about 12\% and the $\mathcal{O}(\alpha_s)$ corrections a further reduction by about 49\%, adding up to a total reduction by about 55\%.
To judge the phenomenological significance of the $\mathcal{O}(v^2)$ corrections in view of the status quo, we should rather compare them to the $\mathcal{O}(\alpha_s)$-corrected results \cite{Yang:2020xkl}, in which case the reduction is as large as $17^{+39}_{-6}$\%.
As for theoretical uncertainties, we observe from Table~\ref{tab:tot} that their absolute sizes are reduced whenever $\mathcal{O}(v^2)$ or $\mathcal{O}(\alpha_s)$ corrections are included.
Contrary to na\"{\i}ve expectations, the relative uncertainty is slightly increased upon inclusion of the $\mathcal{O}(\alpha_s)$ corrections.
This is because of the abnormally large reductions of the central predictions under the influence of the $\mathcal{O}(v^2)$ and $\mathcal{O}(\alpha_s)$ corrections.

Experimentally, $J/\psi$ mesons may be most easily detected and reconstructed through their decays to $e^+e^-$ and $\mu^+\mu^-$ pairs, with a combined branching fraction of $\mathrm{Br}(J/\psi\rightarrow l^+l^-)=12\%$ \cite{ParticleDataGroup:2022pth}.
Dressing the final predictions in Table~\ref{tab:tot} with two factors of $\mathrm{Br}(J/\psi\rightarrow l^+l^-)$ and the respective integrated luminosities $\int \mathrm{d}t\,\mathcal{L}$ per experiment from Table~\ref{Tab:colliders}, we obtain $649^{+82}_{-488}$, $584^{+75}_{-439}$, and $558^{+64}_{-414}$ signal events per year at FCC-ee, CEPC, and CLIC, respectively.

\begin{figure}[htp]
  \begin{tabular}{ccc}
  \includegraphics[width=0.32\textwidth]{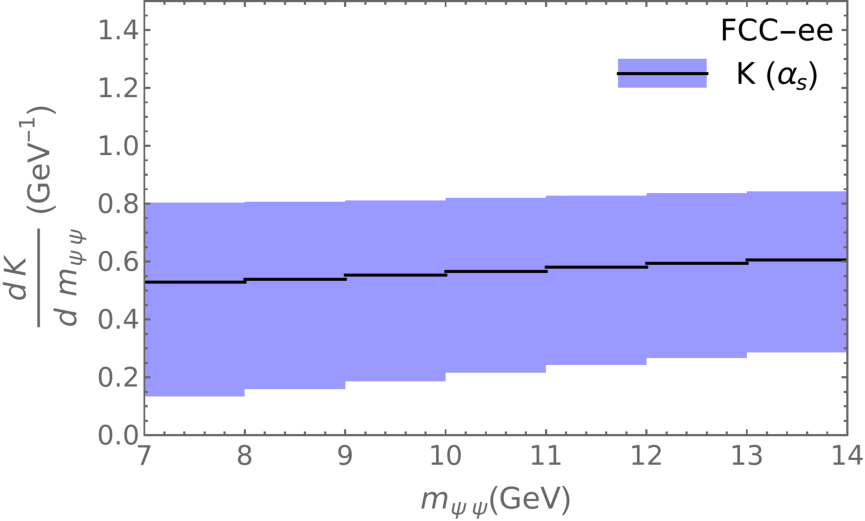}&
  \includegraphics[width=0.32\textwidth]{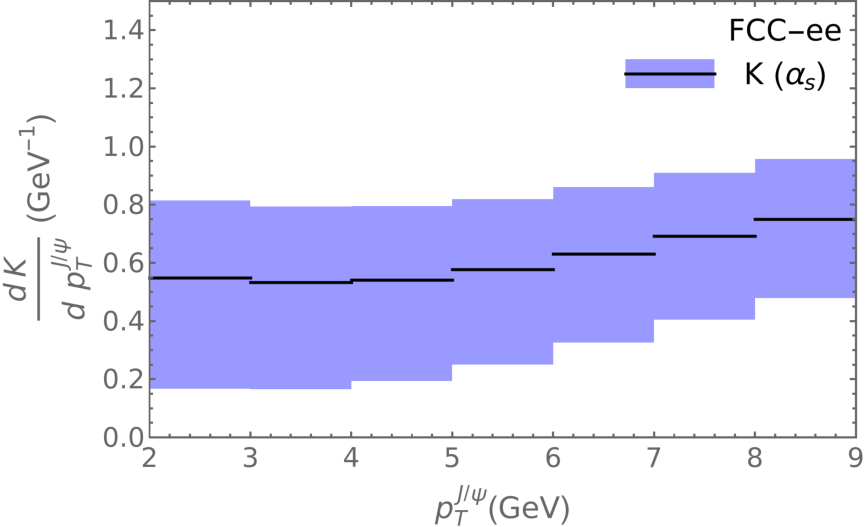}&
  \includegraphics[width=0.32\textwidth]{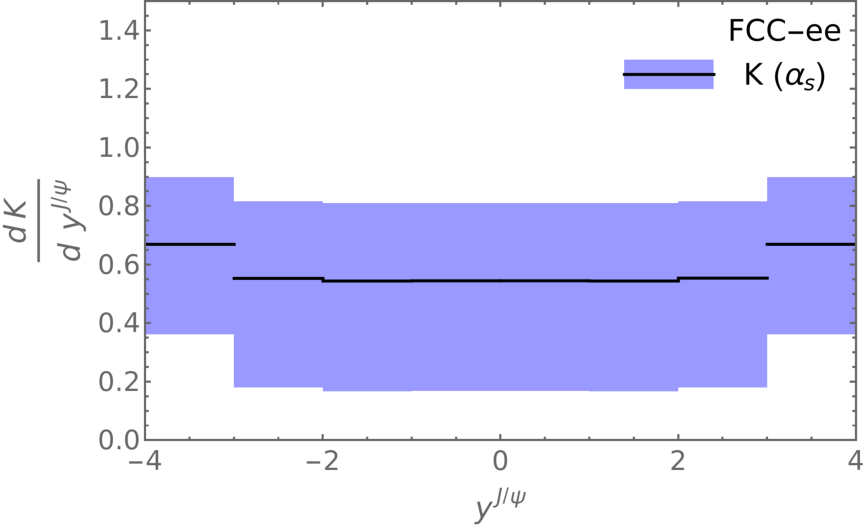}\\
  \includegraphics[width=0.32\textwidth]{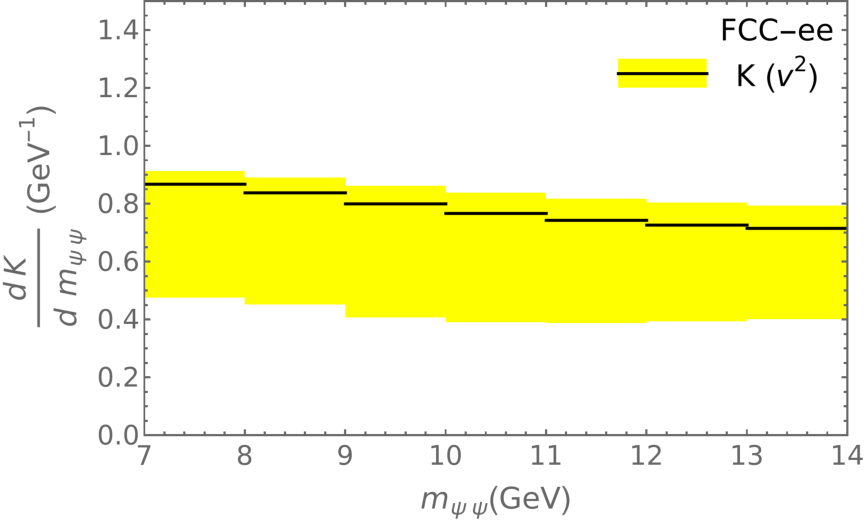}&
  \includegraphics[width=0.32\textwidth]{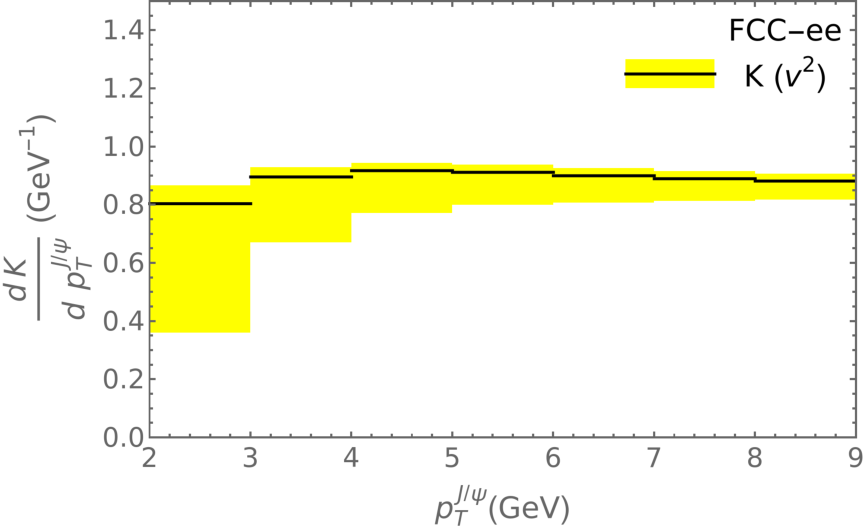}&
  \includegraphics[width=0.32\textwidth]{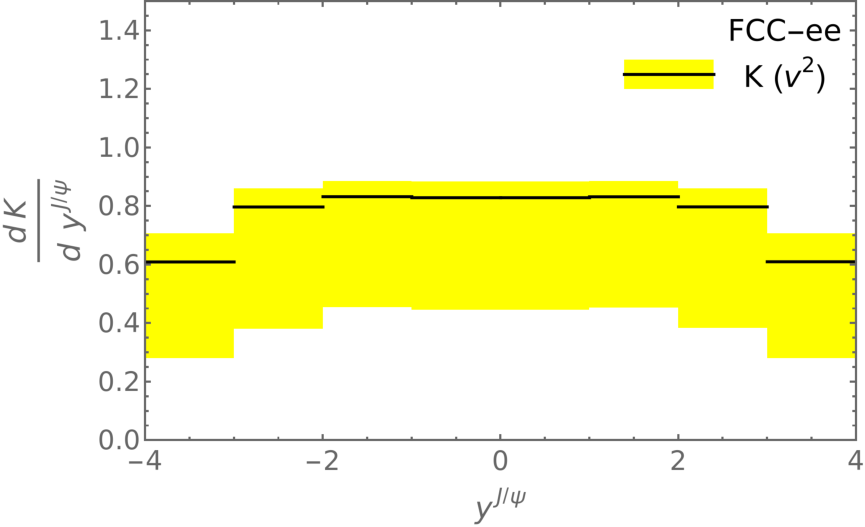}
  \end{tabular}
  \caption{\label{fig:k}%
NRQCD predictions of $m_{\psi \psi}$ (left), $p_T^{J/\psi}$ (center), and $y^{J/\psi}$ (right) distributions of $K_{\alpha_s}$ (top) and $K_{v^2}$ (bottom) at FCC-ee.
    The theoretical uncertainties in $K_{\alpha_s}$ and $K_{v^2}$ are indicated by the shaded blue and yellow bands, respectively.}
\end{figure}

We now turn our attention to cross section distributions in the $J/\psi$ pair invariant mass $m_{\psi\psi}$ and the transverse momentum $p_T^{J/\psi}$ and rapidity $y^{J/\psi}$ of any $J/\psi$ meson.
Specifically, we consider the ranges $7~\mathrm{GeV}\le m_{\psi\psi}\le 14~\mathrm{GeV}$, $2~\mathrm{GeV}\le p_T^{J\psi}\le 9~\mathrm{GeV}$, and $-4\le y^{J/\psi}\le 4$ in bins of unit length, yielding 7, 7, and 8 bins, respectively.

To obtain a more detailed picture of the sizes of corrections and theoretical uncertainties, we study the two consecutive correction factors $K_{\alpha_s}=\mathrm{d}\sigma^{\mathcal{O}(\alpha_s)}/\mathrm{d}\sigma^{\mathrm{LO}}$ and $K_{v^2}=\mathrm{d}\sigma^{\mathcal{O}(\alpha_s,v^2)}/\mathrm{d}\sigma^{\mathcal{O}(\alpha_s)}$.
For this purpose, we may concentrate on the FCC-ee case because the differences with respect to the CEPC and CLIC cases largely cancel out in the ratios.
Our NRQCD predictions for $K_{\alpha_s}$ and $K_{v^2}$ are presented in Fig.~\ref{fig:k}.
We observe from Fig.~\ref{fig:k}, that the central values of $K_{\alpha_s}$ range between 0.55 and 0.75 in the $m_{\psi\psi}$, $p_T^{J/\psi}$, and $y^{J/\psi}$ ranges considered; the $m_{\psi\psi}$ and $|y^{J/\psi}|$ distributions monotonically increase, and so does the $p_T^{J/\psi}$ distribution beyond $p_T^{J/\psi}=4~\mathrm{GeV}$.
On the other hand, Fig.~\ref{fig:k} shows that $K_{v^2}$ ranges from 0.6 to 0.9 and exhibits $m_{\psi\psi}$, $p_T^{J/\psi}$, and $y^{J/\psi}$ line shapes that behave inversely to $K_{\alpha_s}$.
In particular, the relativistic corrections are most important at large values of $m_{\psi\psi}$ and $|y^{J/\psi}|$, and small values of $p_T^{J/\psi}$.
We conclude that the overall correction factor $K=\mathrm{d}\sigma^{\mathcal{O}(\alpha_s,v^2)}/\mathrm{d}\sigma^{\mathrm{LO}}=K_{\alpha_s}K_{v^2}$ exhibits a strongly reduced variation with $m_{\psi\psi}$, $p_T^{J/\psi}$, and $y^{J/\psi}$. 

\begin{figure}[htp]
\begin{tabular}{ccc}
  \includegraphics[width=0.32\textwidth]{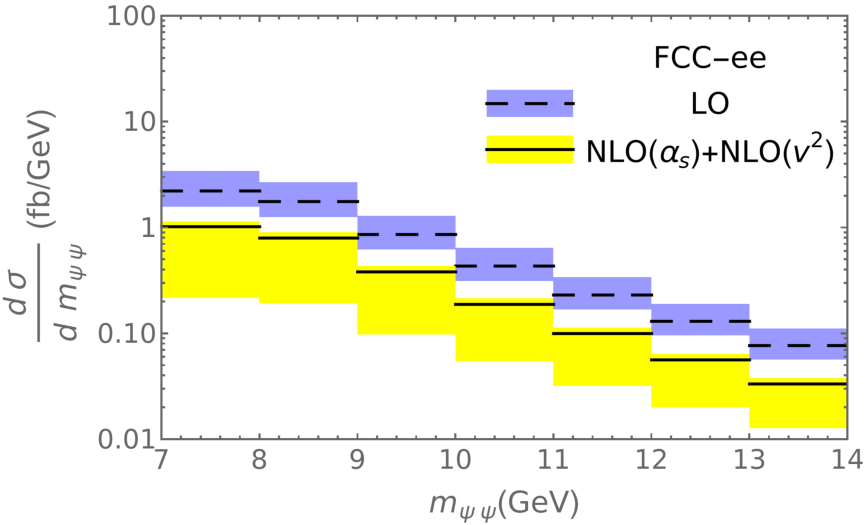}&
  \includegraphics[width=0.32\textwidth]{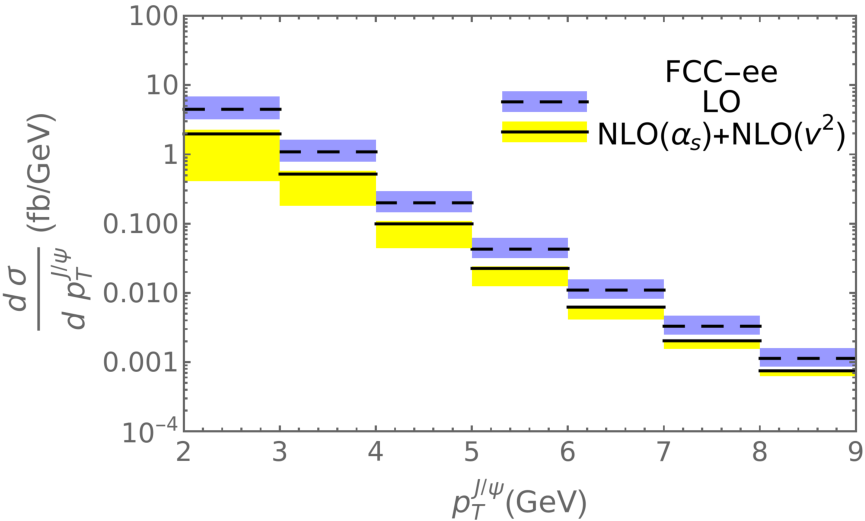}&
  \includegraphics[width=0.32\textwidth]{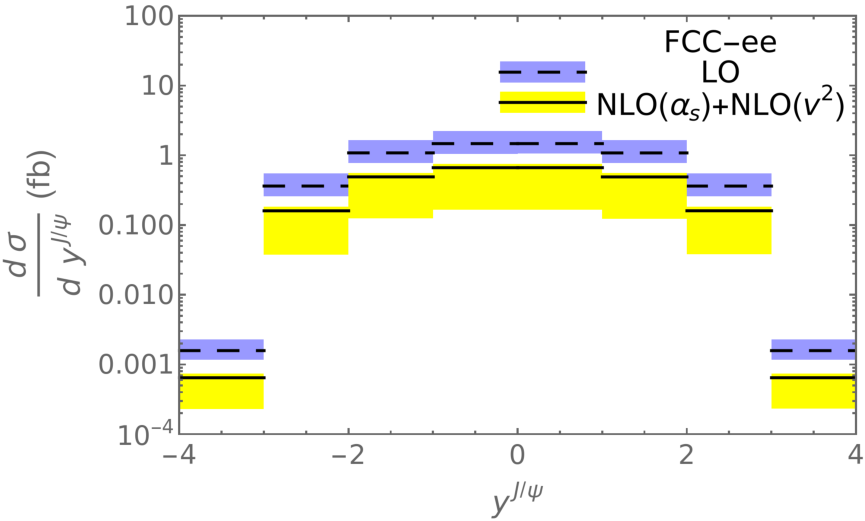}\\
  \includegraphics[width=0.32\textwidth]{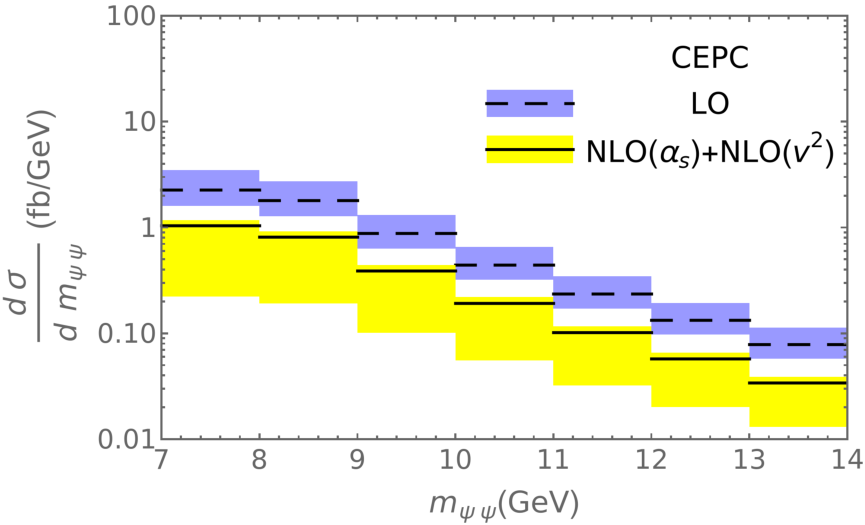}&
  \includegraphics[width=0.32\textwidth]{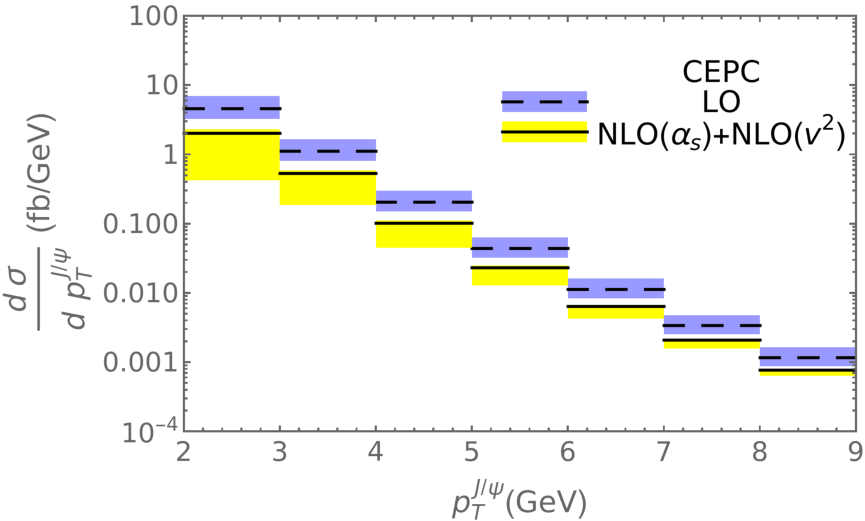}&
  \includegraphics[width=0.32\textwidth]{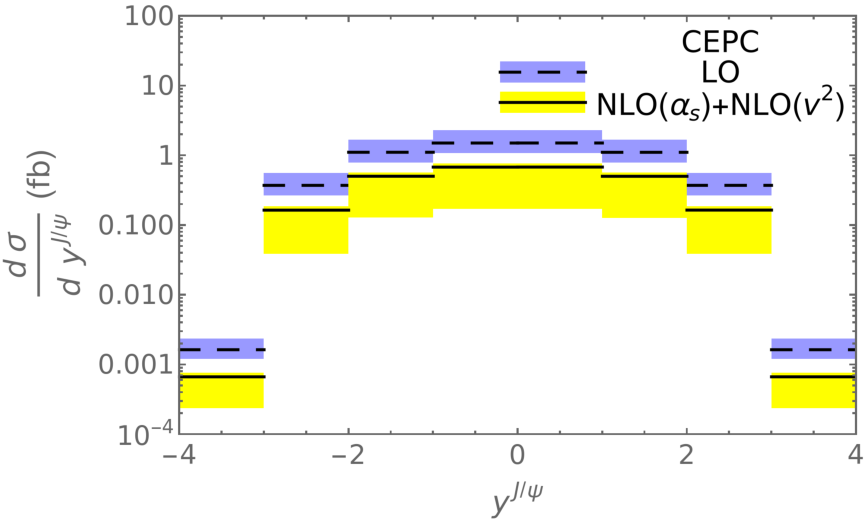}\\
  \includegraphics[width=0.32\textwidth]{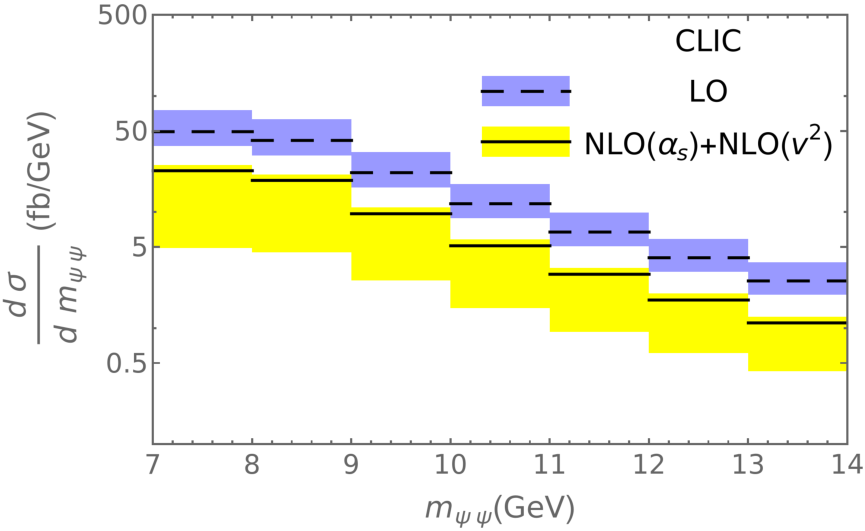}&
  \includegraphics[width=0.32\textwidth]{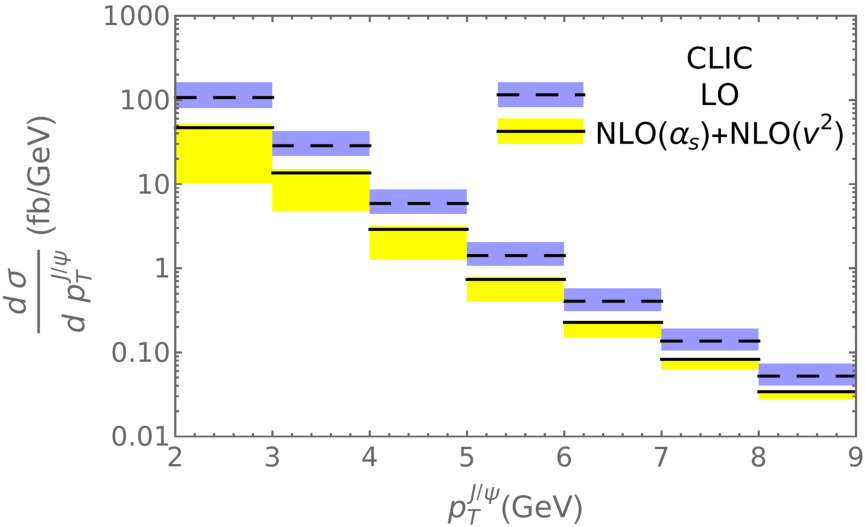}&
  \includegraphics[width=0.32\textwidth]{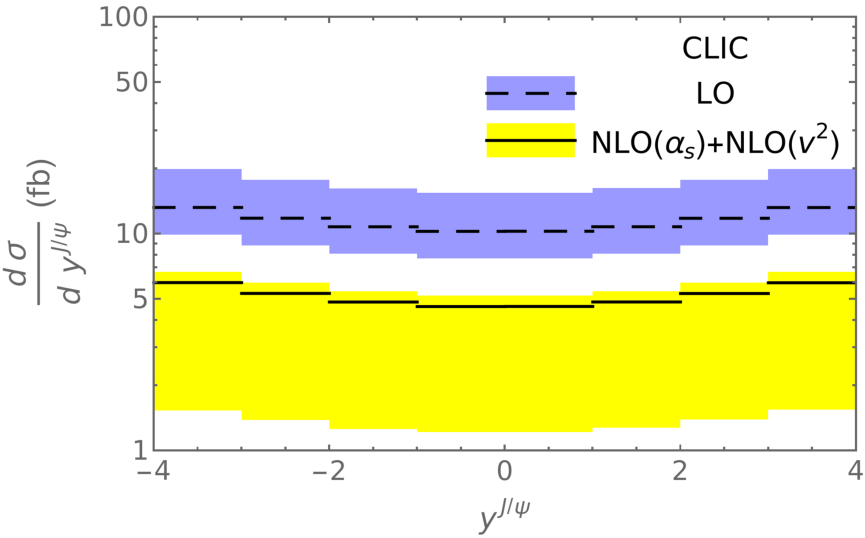}
\end{tabular}
\caption{\label{fig:diff}%
LO and full NLO NRQCD predictions of
$\mathrm{d}\sigma/\mathrm{d}m_{\psi \psi}$ [fb/GeV] (left),
$\mathrm{d}\sigma/\mathrm{d}p_T^{J/\psi}$ [fb/GeV] (center), and
$\mathrm{d}\sigma/\mathrm{d}y^{J/\psi}$ [fb] (right)
for $e^+e^-\to 2J/\psi+X$ via photoproduction
at FCC-ee (top), CEPC (center), and CLIC (bottom).
The theoretical uncertainties at NLO and NLO are indicated by the shaded blue and yellow bands, respectively.}
\end{figure}

In Fig.~\ref{fig:diff}, we present binned $m_{\psi\psi}$, $p_T^{J/\psi}$, and $y^{J/\psi}$ distributions of $e^+e^-\to 2J/\psi+X$ via photoproduction in NRQCD at LO and full NLO, including both $\mathcal{O}(v^2)$ and $\mathcal{O}(\alpha_s)$ corrections, for the FCC-ee, CEPC, and CLIC experimental setups.
We observe from Fig.~\ref{fig:diff} (top row, center column) that the $K$ factors vary only feebly with $m_{\psi\psi}$ and $y^{J/\psi}$ and moderately with $p_T^{J/\psi}$, as expected from the above discussion of Fig.~\ref{fig:k}. In fact, again taking FCC-ee as a representative example, we have $K=045_{-0.37}^{+0.26}$ for the $m_{\psi\psi}$ and $y^{J/\psi}$ distributions, except for bins 1 and 8 of the latter, where $K=0.41_{-0.30}^{+0.23}$.
For the $p_T^{J/\psi}$ distribution, the $K$ factor ranges from $0.44_{-0.38}^{+0.27}$ in the first bin up to $0.66_{-0.27}^{+0.21}$ in the last bin.

\begin{table}
  \caption{\label{tab:diff}
  Numbers of expected signal events per year in each bin of $m_{\psi \psi}$, $p_T^{J/\psi}$ and $y^{J/\psi}$ at FCC-ee (top), CEPC (center), and CLIC (bottom).}
  \begin{ruledtabular}
  \begin{tabular}{ccccccccc}
    bin &1&2&3&4&5&6&7&8\\
    \hline
  $m_{\psi\psi}$&$250^{+30}_{-197}$& $195^{+25}_{-148}$& $93.5^{+12.4}_{-70.0}$& $46.1^{+6.5}_{-32.9}$& $24.4^{+3.4}_{-16.7}$& $13.8^{+1.9}_{-8.9}$& $8.17^{+1.09}_{-5.05}$& -\\
  $p_T$&$486^{+66}_{-386}$& $128^{+14}_{-83}$& $24.4^{+2.3}_{-13.7}$& $5.56^{+0.38}_{-2.49}$& $1.53^{+0.07}_{-0.52}$& $0.501^{+0.010}_{-0.124}$& $0.185^{+0.001}_{-0.032}$& -\\
  $y$&$0.159^{+0.023}_{-0.102}$& $39.4^{+5.3}_{-30.2}$& $121^{+15}_{-90}$& $164^{+21}_{-123}$& $164^{+21}_{-123}$& $121^{+15}_{-90}$& $39.4^{+5.3}_{-30.1}$& $0.159^{+0.023}_{-0.102}$\\
  \hline
  $m_{\psi\psi}$&$225^{+27}_{-177}$& $176^{+23}_{-134}$& $84.2^{+11.0}_{-62.3}$& $41.6^{+5.8}_{-29.6}$& $22.0^{+3.1}_{-15.0}$& $12.4^{+1.7}_{-8.1}$& $7.37^{+0.98}_{-4.55}$\\
  $p_T$&$438^{+59}_{-347}$& $115^{+13}_{-75}$& $22.0^{+2.1}_{-12.3}$& $5.01^{+0.35}_{-2.25}$& $1.38^{+0.06}_{-0.47}$& $0.452^{+0.010}_{-0.112}$& $0.167^{+0.001}_{-0.029}$\\
  $y$&$0.144^{+0.021}_{-0.093}$& $35.5^{+4.8}_{-27.2}$& $109^{+14}_{-81}$& $147^{+19}_{-110}$& $147^{+19}_{-111}$& $109^{+14}_{-81}$& $35.6^{+4.8}_{-27.2}$& $0.144^{+0.021}_{-0.093}$\\
\hline
  $m_{\psi\psi}$&$197^{+22}_{-154}$& $162^{+20}_{-123}$& $83.8^{+10.7}_{-61.6}$& $44.4^{+5.8}_{-31.5}$& $25.2^{+3.3}_{-17.2}$& $15.1^{+2.0}_{-9.9}$& $9.61^{+1.18}_{-5.93}$\\
  $p_T$&$407^{+50}_{-319}$& $118^{+13}_{-77}$& $25.2^{+2.4}_{-14.2}$& $6.39^{+0.46}_{-2.95}$& $1.97^{+0.08}_{-0.69}$& $0.718^{+0.016}_{-0.184}$& $0.296^{+0.002}_{-0.053}$\\
  $y$&$51.5^{+6.2}_{-38.3}$& $45.9^{+5.5}_{-34.0}$& $41.9^{+5.0}_{-31.0}$& $39.9^{+4.8}_{-29.4}$& $40.0^{+4.8}_{-29.5}$& $41.9^{+5.0}_{-30.9}$& $45.9^{+5.5}_{-33.9}$& $51.4^{+6.2}_{-38.1}$
  \end{tabular}
  \end{ruledtabular}
\end{table}

The respective numbers of signal events per year, based on our best predictions, are collected in Table~\ref{tab:diff}.
For all three experimental setups, promising yields are expected in the lower $m_{\psi\psi}$ and $p_T^{J/\psi}$ ranges and in the central $y^{J/\psi}$ region.

\begin{table}
  \caption{\label{tab:mumt}
Full-NLO NRQCD predictions of $\mathrm{d}\sigma/\mathrm{d}p_T$ [fb/GeV] at FCC-ee for $\mu_r=\xi m_T$.}
  \begin{ruledtabular}
  \begin{tabular}{cccccccc}
bin & 1 & 2 & 3 & 4 & 5 & 6 & 7 \\ \hline
  $\xi=1/2$ & $-9.42$ & $-1.90$ & $-0.272$ & $-0.0430$ & $-0.00726$ & $-0.00120 $ & $-0.000124$ \\
  $\xi=1$ & $-0.162$ & 0.0673 & 0.0270 & 0.00974 & 0.00370 & 0.00148 & 0.000620\\
$\xi=2$ & 1.81 & 0.519 & 0.0100 & 0.023 & 0.00647 & 0.00213 & 0.000790
  \end{tabular}
  \end{ruledtabular}
\end{table}

We are faced by considerable theoretical uncertainties in Tables~\ref{tab:tot} and \ref{tab:diff} and Figs.~\ref{fig:k} and \ref{fig:diff}, which manifest themselves in sizable shifts under $\mu_r$ variations.
While we believe that our default choice $\mu_r=\sqrt{s}$ of renormalization scale is most appropriate for the problem at hand, it is instructive to also consider alternatives.
In the case of $J/\psi$ single or associated production, the $J/\psi$ transverse mass $m_T=\sqrt{p_T^2+4m_c^2}$ is often chosen as the default.
This motivates us to explore the choice $\mu_r=\xi m_T$ with $\xi=1/2,1,2$.
We do this taking $\mathrm{d}\sigma/\mathrm{d}p_T$ at FCC-ee as an example and present our full NLO NRQCD predictions in Table~\ref{tab:mumt}.
These turn out to be considerably smaller than our default predictions in Fig.~\ref{fig:diff} and even partly negative, which disqualifies this low scale choice altogether. 

\section{Conclusions}\label{sec:summary}

A tantalizing aspect of the physics potentials of future high-energy, high-luminosity $e^+e^-$ colliders is to shed light on the mechanism underlying the formation of heavy quarkonia.
$J/\psi$ pair production, which has been extensively studied at the LHC, provides a particularly sensitive probe for that.
While the hadroproduction of $J/\psi$ pairs is plagued by sizable, but poorly known DPS contributions \cite{Kom:2011bd}, the latter are absent in $e^+e^-$ collisions.
This motivated us to study prompt $J/\psi$ pair production in two-photon collisions at future $e^+e^-$ colliders considering the FCC-ee \cite{FCC:2018evy}, CEPC \cite{CEPCStudyGroup:2018rmc}, and CLIC \cite{CLICdp:2018cto} experimental setups.
We derived, for the first time, the relativistic corrections of $\mathcal{O}(v^2)$, from both matrix elements and phase space, and provided an independent check of the quantum corrections of $\mathcal{O}(\alpha_s)$ presented in numerical form in Ref.~\cite{Yang:2020xkl}.
We find that the familiar $\mathcal{O}(\alpha_s)$ reduction \cite{Yang:2020xkl} is significantly amplified by the $\mathcal{O}(v^2)$ corrections.
In fact, the $\mathcal{O}(\alpha_s)$-corrected results are typically reduced by 20\%.
Thanks to the extremely high luminosities envisaged at the future $e^+e^-$ facilities, promising signal yields may still be expected.
Assuming the $J/\psi$ mesons to be reconstructed via their $e^+e^-$ and $\mu^+\mu^-$ decays, we predict $649^{+82}_{-488}$, $584^{+75}_{-439}$, and $558^{+64}_{-414}$ signal events per year at the FCC-ee, CEPC, and CLIC, respectively.
The large theoretical uncertainties are due to the lack of knowledge of higher-order corrections and determined by variations of the renormalization scale as usual.
As a by-product, our results can easily be applied to double $J/\psi$ production in ultra-peripheral collisions at hadron colliders. 

\begin{acknowledgments}

  We thank Cong-Feng Qiao for helpful communications enabling a comparison with the results of Ref.~\cite{Yang:2020xkl}.
  This work was supported in part by the German Research Foundation DFG through Research Unit FOR~2926 ``Next Generation Perturbative QCD for Hadron Structure: 
  Preparing for the Electron-Ion Collider" under Grant No.~409651613.
  The work of X.B.J. was supported in part by National Natural Science Foundation of China under Grant No.~12061131006.
  The work of R.L. was supported in part by National Natural Science Foundation of China under Grant Nos.~U1832160 and 12075177.

\end{acknowledgments}

\end{document}